\documentclass[showpacs,amsmath,amssymb,twocolumn,pra,superscriptaddress,notitlepage]{revtex4-1}
\usepackage{graphicx}				
\usepackage{amssymb,amsmath}
\usepackage{physics}
\usepackage{subfigure}
\usepackage{hyperref} 
\usepackage{ulem}
\usepackage{color}

\begin{document}
\title{Quantum sensing to suppress systematic errors with measurements
 in the Zeno regime
}

\author{Alisa Shimada}
\affiliation{Research Center for Emerging Computing Technologies, National institute of Advanced Industrial Science and Technology (AIST), Central2, 1-1-1 Umezono, Tsukuba, Ibaraki 305-8568, Japan}
\author{Hideaki Hakoshima}
\affiliation{Research Center for Emerging Computing Technologies, National institute of Advanced Industrial Science and Technology (AIST), Central2, 1-1-1 Umezono, Tsukuba, Ibaraki 305-8568, Japan}

\author{Suguru Endo}
\affiliation{NTT Secure Platform Laboratories, NTT Corporation, Musashino 180-8585, Japan}

\author{Kaoru Yamamoto}
\affiliation{NTT Secure Platform Laboratories, NTT Corporation, Musashino 180-8585, Japan}

\author{Yuichiro Matsuzaki}
\email{matsuzaki.yuichiro@aist.go.jp}
\affiliation{Research Center for Emerging Computing Technologies, National institute of Advanced Industrial Science and Technology (AIST), Central2, 1-1-1 Umezono, Tsukuba, Ibaraki 305-8568, Japan}

\begin{abstract}
Quantum magnetic field sensing is an important technology for material science and biology. Although experimental imperfections affect the sensitivity, repetitions of the measurements decrease the estimation uncertainty by a square root of the total number of measurements 
if there are only statistical errors.
However, it is difficult to precisely characterize the coherence time of the system because it fluctuates in time in realistic conditions, which induces systematic errors. In this case, due to residual bias of the measured values, estimation uncertainty cannot be lowered than a finite value even in the limit of the infinite number of measurements. On the basis of the fact that the decoherence dynamics in the so-called Zeno regime are not significant compared to other regimes,  we propose a novel but very simple protocol to use measurements in the Zeno regime for reducing systematic errors. Our scheme allows the estimation uncertainty $\delta ^2 \omega$ to scale as $L^{1/4}$ where $L$ denotes the number of measurements even when we cannot precisely characterize the coherence time.
\end{abstract}
\maketitle

\section{Introduction}
Measurements of small magnetic fields is an essential technique in several fields such as biology and material science. Recently, quantum systems have been used for sensitive magnetic field sensing. External magnetic fields shift the resonant frequency of the quantum system, and measurements of a phase shift allow us to detect magnetic fields via the frequency shift \cite{degen2017quantum}. Nitrogen vacancies in diamond \cite{taylor2008high,maze2008nanoscale,balasubramanian2008nanoscale},  superconducting flux qubits \cite{bal2012ultrasensitive,toida2019electron,budoyo2020electron}, and optically pumped atoms \cite{kominis2003subfemtotesla}
have been used for such quantum magnetic field sensing.

Experimental imperfections such as decoherence typically deteriorate the sensitivity \cite{degen2017quantum}. For the detection of the small magnetic fields, it is essential to decrease the estimation uncertainty. When experimental parameters such as g factor and coherence time of the qubits are known, we can estimate the magnetic fields from the experimental results without bias.
In this case, the estimation uncertainty decreases as the total number of measurements increases, because there are only statistical errors.

However, in the actual experiments, the coherence time of the system generally fluctuates in time \cite{yan2016flux,abdurakhimov2019long}, which makes it difficult to know the accurate value of the coherence time. In such a case, the lack of 
knowledge of the coherence time induces systematic errors in the estimation. Under the effect of the systematic errors, the estimation uncertainty is lower bounded by a specific value even in the limit of the large number of measurements \cite{wolf2015subpicotesla,budoyo2020electron}. This could be a
problem to measure small magnetic fields.

On the other hand,  quantum mechanics predicts that 
frequent measurements with short time intervals can suppress the evolution of quantum states \cite{misra1977zeno,facchi2008quantum}. In the so-called Zeno-regime,  `survival probability' that the state remains in the initial state after a short time $t$ scales as $P_{\rm{suv}}\simeq 1-\Gamma ^2 t^2$, where $\Gamma$ denotes a decay rate \cite{schulman1994characteristic,nakazato1996temporal}  \cite{erez2004correcting,facchi2008quantum,nakazato1996temporal,koshino2005quantum}. This phenomenon has been experimentally demonstrated
in several systems \cite{itano1990quantum,kwiat1995interaction,fischer2001observation,streed2006continuous,helmer2009quantum,wolters2013quantum,kakuyanagi2015observation,kondo2016using,kalb2016experimental}.

Here, we propose to use measurements in the Zeno regime for quantum sensing
to suppress the systematic errors when we cannot characterize the precise value of the coherence time. In the Zeno regime, the system coupled with the environment shows a quadratic decay, and the decay dynamics can be much slower for a short time region than the other regions. This means that the qubit state is not significantly affected by the decoherence in the short time regime. We show that, by performing the measurements in the Zeno regime, we can suppress the effect of the systematic errors induced by the lack of 
knowledge of the coherence time. 
Moreover, in our scheme,
the estimation uncertainty
can be arbitrarily small by increasing the number of  measurements, even
when we do not know the precise value of the coherence time.
Although the measurements in the Zeno regime was discussed in entanglement enhanced sensing \cite{Matsuzaki2011NMmetro,ChinNM2012PRL,Zeno2015}, we firstly utilize this concept to suppress the systematic errors.

\section{quantum sensing}
Let us review the standard quantum sensing, focusing on the case of a qubit.
We explain two cases, without and with systematic errors.

\subsection{Quantum sensing without systematic errors}
First, we review the sensing without systematic errors. 
Here, for simplicity, we ignore any experimental imperfection such as decoherence. 
We consider the Hamiltonian as follows
\begin{eqnarray}
H=\sum _{j=1}^L \frac{\omega}{2} \hat{\sigma}^{(j)}_z,
\end{eqnarray}
where $\omega$ denotes a frequency of the qubit and 
$\hat{\sigma}_z$
denote a Pauli matrix spanned by $|1\rangle $ and $|-1\rangle$. We assume that $\omega$ has a linear dependence on
applied magnetic fields $B$. This means that, if we know the value of the $\omega$, we can determine the value of the magnetic fields.
The sensing protocol can be performed as following.
First, we prepare the state of $\bigotimes _{j=1}^L|+\rangle _j=\frac{1}{\sqrt{2}}(|1\rangle _j+|-1\rangle _j)$.
Second, let the state evolve by the Hamiltonian to obtain $\bigotimes _{j=1}^L|\psi(t) \rangle_j $
where $|\psi(t) \rangle_j=\frac{1}{\sqrt{2}}(|1\rangle _j+e^{i\omega t}|-1\rangle _j)$. Third, perform a projective measurement with $\hat{\mathcal{P}}_j=(\openone +\hat{\sigma }^{(j)}_y)/2$ for $j=1,2,\cdots L$
where the probability to have this projection is $P'=(1+\sin \omega t)/2\simeq (1+\omega t)/2$ for weak magnetic fields with $|\omega| t \ll 1$.
Fourth, repeat these three steps $N$ times.
Finally, we obtain $L$ measurement results $\{s_m\}_{m=1}^{L}$
from $L$ qubits, and estimate the frequency $\omega$ based on the measurement results.
More specifically, we obtain the estimate value as $\omega _{\rm{est}}=(2(\sum _{m=1}^{L} \frac{s_m}{LN})-1)/t$.
We can calculate the uncertainty of the estimation 
 as follows.
\begin{eqnarray}
\delta ^2\omega &=& \overline{(\omega -\omega _{\rm{est}})^2} \nonumber \\
&=&\frac{P'(1-P')}{|\frac{dP'}{d\omega }|^2LN}\nonumber \\
&=&\frac{1}{t^2LN}
\end{eqnarray}
 where the overline denotes the statistical average.
 In our paper, we basically consider a case of $N=1$.
 The uncertainty scales as $\delta ^2 \omega =\Theta(L^{-2})$, and so we can decrease the uncertainty by increasing the number of qubits 

\subsection{Quantum sensing with systematic errors}
We review the sensing with systematic errors. 
Suppose that there are some unknown experimental imperfections such as  systematic errors, and the actual probability $P'$ to have a projection of  $\hat{\mathcal{P}}_j=(\openone +\hat{\sigma }^{(j)}_y)/2$ for $j=1,2,\cdots L$
is different from what an experimentalist expected to be true.
Let us define such a wrong (true) probability 
as $P$ ($P'$). We assume that these probabilities
have a linear
dependence on $\omega$ such as $P=x + y \omega $ and $P'=x'+y'\omega $.
In this case, the uncertainty of the estimation for a small $\omega$ 
can be calculated as follows \cite{sugiyama2015precision,takeuchi2019quantum}.
\begin{eqnarray}
\delta ^2 \omega = \frac{1}{y^2}\left[\frac{P'(1-P')}{L}+(x-x')^2\right]\label{takeuchi}
\end{eqnarray}
It is worth mentioning that, even in a limit of large $L$, the uncertainty has a finite non-zero value as follows: 
\begin{eqnarray}
\lim _{L \rightarrow \infty}\delta ^2 \omega =\frac{(x-x')^2}{y^2}.
\end{eqnarray}
This finite value comes from systematic errors due to the lack of 
knowledge of the coherence time.
The goal of this paper is to suppress such systematic errors.

\section{Quantum sensing with measurements in Zeno regime}
In this section, we explain our scheme to suppress the systematic errors with measurements in Zeno regime.

\subsection{Setup}
Let us explain the setup of our scheme.
We consider a three level system (or a spin-1 system) , which describes
an NV center in diamond \cite{taylor2008high,maze2008nanoscale,balasubramanian2008nanoscale}
or a capacitively shunted flux qubit \cite{you2007low,yan2016flux}.
Although there are three levels such as $|1\rangle $, $|-1\rangle $, and $|0\rangle $, we mainly use $|1\rangle $ and $|-1\rangle $ for magnetic field sensing.
However, due to an energy relaxation, we have unwanted population of the state $|0\rangle $.
Throughout this paper, although this system is not a qubit because it has a small population of the third level, we call this a qubit because we mainly
use $|1\rangle $ and $|-1\rangle $.
In the magnetic field sensing scheme, there is a state preparation step to have an initial state of $\bigotimes _j^L|+\rangle _j=\bigotimes _j^L\frac{1}{\sqrt{2}}(|1\rangle _j+|-1\rangle _j)$, an exposure step to interact the spin-1 systems with target magnetic fields, and a readout step to measure the state.
We assume that the state preparation time and readout time is much longer than the exposure time.
After interacting the $L$ qubits
with the magnetic fields for sensing, we have the following state.
\begin{eqnarray}
\rho &=& \bigotimes _{j=1}^L \rho _j \label{quadstate}
\nonumber \\
\rho _j&=&(1-\epsilon )|\psi (t)\rangle _j \langle \psi(t)|_j+\epsilon  |0\rangle_j \langle 0|_j
\label{rhostate}
 \\
|\psi (t)\rangle _j&=& \frac{1}{\sqrt{2}}(|1\rangle_j +e^{i\omega t}|-1\rangle _j
\end{eqnarray}
where $\epsilon$ denotes an error rate due to the energy relaxation.
Since every unstable system shows a quadratic decay for a short time scale
\cite{schulman1994characteristic,nakazato1996temporal}, the error can be approximately described by  $1-\epsilon \simeq 1- (t/T'_1)^2 $, where $T'_1$ denotes a energy relaxation time
as long as we are interested in a short time scale.
We assume that the experimentalists do not know the precise value of $T'_1$.
In the experiment, the energy relaxation time can 
fluctuate
in time
\cite{yan2016flux,abdurakhimov2019long}. In this case, we could not estimate the exact value of $T'_1$ because it changes before a precise estimation of $T'_1$.
In our paper, we assume that, $T'_1$ is constant during the interaction between the spin-1 system and magnetic fields. However, it is worth mentioning that
$T'_1$ could change if we consider a longer time scale such as an initialize time, a readout time, or a repetition time (when we have $N\geq 2$), which makes it difficult to estimate the exact value of the 
$T_1'$.

\subsection{Uncertainty of the estimation for a Gaussian decay}
Let us calculate the uncertainty of the estimation when we have a Gaussian decay such as 
$\epsilon =1- e^{-(t/T'_1)^2} $.
We perform 
projective measurement with $\hat{\mathcal{P}}_j=(\openone +\hat{\sigma }^{(j)}_y)/2$ for $j=1,2,\cdots L$ on the state described by Eq.~(\ref{rhostate}). 
Here, $\hat{\sigma }^{(j)}_y$ is the Pauli matrix spanned by $|1\rangle $ and $|-1\rangle $.
The probability to obtain this projection is as follows.
\begin{eqnarray}
P'\simeq \frac{e^{-(t/T'_1)^2}}{2}(1+ \omega t)
\end{eqnarray}
where we assume $|\omega t|\ll 1$. 
Since we do not know the precise value of $T'_1$, we use $T_1 (\neq T_1')$ to estimate the projection probability.
This means that we consider the projection probability as $P\simeq \frac{e^{-(t/T_1)^2}}{2}(1+ \omega t)$, which is different from the true probability $P'$.
In this case, the imperfect knowledge of the energy relaxation time induce the systematic errors. We can calculate the estimation uncertainty 
from Eq.~(\ref{takeuchi}) as follows.
\begin{eqnarray}
\delta ^2 \omega &\simeq&\frac{e^{2(t/T_1)^2}}{t^2}\Bigl[\frac{e^{-(t/T'_1)^2}(2-e^{-(t/T'_1)^2})}{L}\Bigr.\nonumber \\
&+& \Bigl.(e^{-(t/T_1)^2}-e^{-(t/T'_1)^2})^2\Bigr],\label{gaussw}
\end{eqnarray}
where the first term comes from the statistical error while the second term comes from the systematic errors due to the lack of 
the
knowledge of the coherence time.

Here, we try to minimize the uncertainty by optimizing the interaction time $t$.
In the conventional strategy, 
since
we believe that $T_1 \simeq T_1'$, 
we obtain the following 'wrong' uncertainty by substituting $T_1'=T_1$ in Eq.~(\ref{gaussw}):
\begin{eqnarray}
\delta ^2 \omega _{\rm{est}}= \frac{2e^{(t/T_1)^2}-1}{(t/T_1)^2LT_1^2}.
\end{eqnarray}
Actually, we can minimize $\delta ^2 \omega _{\rm{est}}$ with $t= c T_1$ and $c\simeq 0.876$.
However, since the actual 
coherence time $T_1'$ is different from $T_1$, the optimal $t$ obtained above induces the bias. To check this,
we calculate the 'actual' uncertainty under $T_1'\neq T_1$ for
$t= c T_1$ 
using
Eq.~(\ref{gaussw}) 
and obtain
\begin{eqnarray}
\delta ^2 \omega _{\rm{conv}}&\simeq &
\frac{e^{2c^2}}{t^2}\Bigl[\frac{e^{-(cT_1/T_1')^2}(2-e^{-(cT_1/T_1')^2})}{L}\Bigr.\nonumber \\
&+& \Bigl.(e^{-c^2}-e^{-(cT_1/T_1')^2})^2\Bigr] .
\end{eqnarray}
We can confirm that
even in the limit of large $L$, the uncertainty approaches to $\lim _{L\rightarrow \infty }\delta ^2 \omega _{\rm{conv}} =(e^{-c^2}-e^{-(cT_1/T_1')^2})^2)$, and so the uncertainty is lower bounded due to the systematic errors. This could be a serious problem to use the quantum sensor for practical purposes.

\begin{figure}
\centering
\includegraphics[width=1.0\columnwidth]{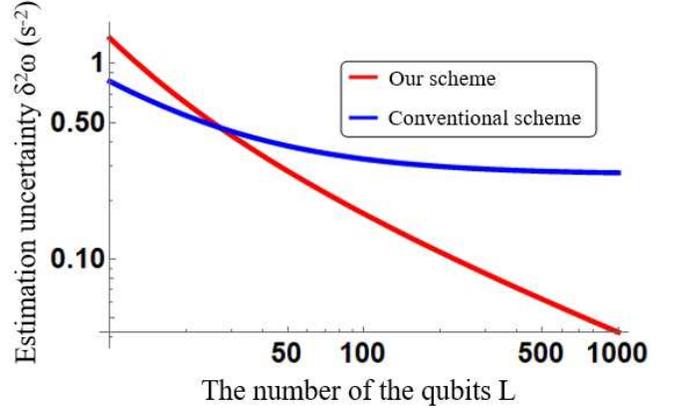}
\caption{The uncertainty of the estimation against the number of qubits with systematic errors.
We set $T_1=1$ $s$, $T_1'=1.4$ $s$, and $t =2L^{-1/4}$ $s$. The estimation uncertainty of our scheme becomes smaller than that of the conventional scheme for $L\geqq 26$.}
\label{fig:gauss}
\end{figure}
On the other hand, in our strategy, we set the interaction time $t$ in the Zeno regime to suppress the systematic errors. We consider a case with $t/T'_1 \ll 1$. To satisfy this condition, we remark that it is not necessary to know the exact value of $T_1'$, but just necessary to know the order of $T_1'$.
In this case, we can perform a Taylor expansion on the estimation uncertainty
for $t/T'_1 \ll 1$, and we obtain the following.
\begin{eqnarray}
\delta ^2 \omega &\simeq&\frac{1}{Lt^2}+\frac{1}{t^2} \left[\left(\frac{t}{T_1}\right)^2-\left(\frac{t}{T'_1}\right)^2\right]^2\nonumber \\
&\geq&\frac{2}{\sqrt{L}}\left| \left(\frac{1}{T_1}\right)^2-\left(\frac{1}{T'_1}\right)^2\right|
\label{newgaussw}
\end{eqnarray}
where we use an inequality of 
arithmetic 
and geometric means. The interaction time that minimize $\delta^2\omega$ is given as follows:
\begin{eqnarray}
\frac{1}{Lt^2}&=&\frac{1}{t^2} \left[\left(\frac{t}{T_1}\right)^2-\left(\frac{t}{T'_1}\right)^2\right]^2\nonumber \\
\Leftrightarrow t&=&t_{\rm{opt}}=L^{-1/4}\left| \left(\frac{1}{T_1}\right)^2-\left(\frac{1}{T'_1}\right)^2\right|^{-1/2} .
\end{eqnarray}
It is noteworthy that, since we do not know the precise value of $T'_1$, we cannot choose the 
optimal
time $t_{\rm{opt}}=L^{-1/4}|(1/T_1)^2-(1/T'_1)^2|^{-1/2}$.
Instead, we can take $t=\tau L^{-1/4}$ where $\tau$ can be determined by a rough estimation of $T_1'$. 
Therefore, by taking $t=\Theta(L^{-1/4})$, we obtain $\delta ^2\omega =\Theta(L^{-1/2})$ regardless 
of
the choice of $\tau$.
This means that, by increasing the number of qubits, we can decrease the estimation uncertainty as small as we want, which is in stark contrast to the conventional scheme to have a lower bound of the estimation uncertainty.
Note that, although such measurements in the Zeno regime were utilized
in entanglement enhanced sensing 
\cite{Matsuzaki2011NMmetro,ChinNM2012PRL,Zeno2015}, we firstly utilize this concept to suppress the systematic errors of the quantum sensing.

We plot the uncertainty of the estimation with the systematic errors
to
compare the uncertainty of our scheme with that of the conventional scheme, as shown in Fig.~\ref{fig:gauss}.
The uncertainty of the conventional scheme approaches to a finite non-zero value, while the uncertainty of our scheme decreases as we increase the number of qubits.
In our scheme, the statistical error is larger than that of the conventional scheme due to the short interaction time between the qubits and magnetic fields. So, for the small number of the qubits where the statistical error is more dominant than the systematic errors, the uncertainty of our scheme is larger than that of the conventional scheme.
However, as we increase the number of qubits, the systematic errors become more dominant than the statistical error, and our scheme shows a better performance than the conventional scheme.

\subsection{Uncertainty of the estimation for a realistic decay}
Here, we calculate the estimation uncertainty for a more realistic decay.
We consider a spin-boson model with a Lorentzian form factor, and this is a typical noise model when the spin-1 system is coupled with a leaky cavity \cite{koshino2005quantum}.
(See the details in Appendix \ref{general}). 
The error rate can be described as follows:
\begin{eqnarray}
\epsilon&=&1-|f(t)|^2\nonumber \\
f(t)&=&\frac{1+\sqrt{1-2\gamma /\Delta}
 }{2e^{i \lambda _1 t}\sqrt{1-2\gamma /\Delta}}
 -
 \frac{1-\sqrt{1-2\gamma /\Delta}
 }{2e^{i \lambda _2 t}\sqrt{1-2\gamma /\Delta}}\ \ \ \ \label{ffunction}
\end{eqnarray}
where $\gamma$ denotes the strength of the noise, 
and $\Delta$ denotes the linewidth of the form factor.
Also, $\lambda _1$ and $\lambda _2$ are the solutions of the following equation:
\begin{eqnarray}
\lambda (\lambda +i\Delta) - \gamma \Delta/2=0 .
\end{eqnarray}
We can find the coherence time of the system as $\tilde{T}'_1=\sqrt{2/(\gamma \Delta)}$ by approximating the error rate as $\epsilon \simeq 1- \frac{\gamma \Delta t^2}{2}$.
The $f(t)$ depends on $\gamma$ and $\tilde{T}'_1$. We assume that, while we know the value of $\gamma$, we do not know the exact value of $\tilde{T}'_1$, and we have $\tilde{T}_1$ as an expected coherence time where $\tilde{T}'_1\neq \tilde{T}_1$. By substituting $\Delta = \frac{2}{\gamma (\tilde{T_1})^2}$ with Eq.~(\ref{ffunction}), we obtain $f_{\rm{est}}(t)$ that we expect, while the true value is $f_{\rm{true}}(t)$ where $\Delta = \frac{2}{\gamma (\tilde{T'_1})^2}$ is substituted with Eq.~(\ref{ffunction}).

Let us calculate the uncertainty of the estimation when we have the decay such as
$\epsilon =1-|f(t)|^2$.
We perform 
projective measurement with $\hat{\mathcal{P}}_j=(\openone +\hat{\sigma }^{(j)}_y)/2$ for $j=1,2,\cdots L$ 
on the state 
described
by Eq.~(\ref{rhostate}). 
The probability to obtain this projection is as follows:
\begin{eqnarray}
P\simeq \frac{1-|f(t)|^2}{2}(1+ \omega t),
\end{eqnarray}
where we assume $|\omega t|\ll 1$. 
If the true value of the coherence time is given, we have 
$P'\simeq \frac{1-|f_{\rm{true}}(t)|^2}{2}(1+ \omega t)$, while we have
$P\simeq \frac{1-|f_{\rm{est}}(t)|^2}{2}(1+ \omega t)$ when we have $f_{\rm{est}}(t)$.
By using Eq.~(\ref{takeuchi}), we obtain
\begin{eqnarray}
\delta ^2 \omega &=& \frac{1}{t^2(1-|f_{\rm{est}}(t)|^2)^2}\Bigl[\frac{(1-|f_{\rm{true}}(t)|^2)(1+|f_{\rm{true}}(t)|^2)}{L}
\Bigr.\nonumber \\
&+&\Bigl.(|f_{\rm{est}}(t)|^2-|f_{\rm{true}}(t)|^2)^2\Bigr]\label{genedw}
\end{eqnarray}
As long as we are interested in a short time region, we can obtain an approximate form of the error rate such as $1-|f_{\rm{true}}(t)|^2 \simeq 1- t^2/(\tilde{T}'_1)^2\simeq e^{- (t/\tilde{T}'_1)^2}$ and $1-|f_{\rm{est}}(t)|^2 \simeq e^{- (t/\tilde{T}_1)^2}$.
In this case, the uncertainty of Eq.~(\ref{genedw})
has the same form as Eq.~(\ref{gaussw}), and 
we can adopt the same strategy as the case of the Gaussian decay.
Therefore, we can take 
$t=\Theta(L^{-1/4})$ to obtain the scaling of $\delta ^2\omega =\Theta(L^{-1/2})$.

\begin{figure}
\centering
\includegraphics[width=1.0\columnwidth]{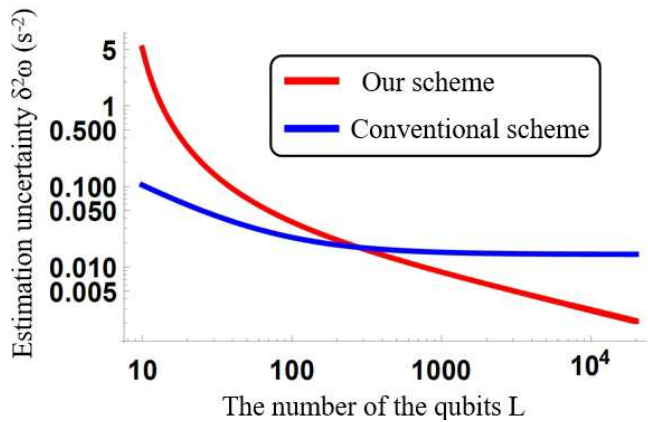}
\caption{The uncertainty of the estimation against the number of 
qubits
with systematic errors.
We set $\tilde{T}_1=2$ $s$, $\tilde{T}'_1=2.4$ $s$, $\gamma =1$ $s^{-1}$, and $t=2L^{-1/4}$ $s$. The estimation uncertainty of our scheme becomes smaller than that of the conventional scheme for $L\geqq 279$.}
\label{fig:koshino}
\end{figure}

To quantify the performance of our scheme, we plot the uncertainty of the estimation of our scheme, and also compare it with that of the conventional scheme in Fig.~\ref{fig:koshino}.
In the conventional scheme, similar to the Gaussian decay case, we substitute $\tilde{T}'_1=\tilde{T}_1$ in
Eq.~(\ref{genedw}), and obtain $\delta ^2 \omega_{\rm{conv}}$. After we fix the values of $\gamma$, $\tilde{T}_1$, and $\tilde{T}'_1$, we minimize $\delta ^2 \omega_{\rm{conv}}$ by choosing the optimal $t$, and use this $t$ for the plot. On the other hand, in our scheme, we take $t=\tau L^{-1/4}$.
Similar to the case of the Gaussian decay, the statistical error is relevant for the small number of the qubits, and the uncertainty of our scheme is larger than that of the conventional scheme.
However, the uncertainty of the estimation of the conventional scheme has a lower bound even in the large $L$ due to the systematic errors. On the other hand, in our scheme, the uncertainty decreases as we increase the number of qubits. Therefore, for the large number of the qubits, our scheme shows better performance than the conventional one.

\section{Conclusion}
In conclusion, we propose a scheme to use measurements in the Zeno regime for the suppression of systematic errors in quantum magnetic field sensing. When we do not know a precise value of the coherence time of the quantum system, such a lack of 
knowledge induces the systematic errors for quantum sensing. In this case, the uncertainty of the estimation of the target magnetic field is lower bounded even in the limit of the infinite number of the repetitions of the measurements.
To suppress such systematic errors, we utilized the fact that every decoherence process shows a quadratic decay in a short time region in the Zeno regime where the effect of the decoherence does not significantly affect the dynamics. We show that, by taking the interaction time (between the system and magnetic fields) as $\Theta (L^{-1/4})$, we obtain a scaling of $\delta ^2 \omega =\Theta (L^{-1/2})$ where $L$ is the number of measurements. 
Therefore, by increasing the number of 
the
repetition, we can decrease the uncertainty as small as we want even under the effect of the systematic errors due to the imperfect knowledge of the coherence time.
Our results pave the way for the practically useful quantum magnetic fields sensors.

\begin{acknowledgments}
We are grateful to Shiro Kawabata for useful discussions.
This work was supported by Leading Initiative for Excellent Young Researchers MEXT Japan and JST presto (Grant No. JPMJPR1919) Japan.
\end{acknowledgments}

\appendix

\section{Decay dynamics under the effect of the decoherence}\label{general}
Here, we discuss the decay dynamics of the three level system
under the effect of the decoherence. 
Especially, we consider an NV center for the three level system.

The Hamiltonian of the NV center coupled with an environment 
is described as follows.
\begin{eqnarray}
H&=&H_{nv}+H_{I}+H_{E} \nonumber \\
 H_{nv}&=&D_0 \hat{S}_z^2 + g\mu  _bB \hat{S}_z \nonumber \\
 H_{I}&=&  \hat{S}_x \int _{\mu} g_{\mu}(\hat{b}_{\mu} +\hat{b}_{\mu}^{\dagger})d \mu \nonumber \\
 H_E&=& \int _{\mu} \omega _{\mu} \hat{b}^{\dagger}_{\mu} \hat{b}_{\mu} d\mu
\end{eqnarray}
where $D_0$ denotes a zero field splitting, 
$g$ denotes a g factor, $\mu _b$ denotes a Bohr magneton, $g_\mu$ denotes the coupling strength 
between the NV center and environment, $\omega _\mu$ denotes the frequency of the environment,
$\hat{S}_x$ ($\hat{S}_y$) denote the spin $1$ operator for $x$ ($y$), and $\hat{b}_\mu$ denotes 
an
annihilation (creation) operator of the environmental mode.
In the interaction picture,
we have
\begin{eqnarray}
 H_I(t)=
  \int _{\mu}g_{\mu} (|B\rangle \langle 0| \hat{b}_{\mu}e^{-i \Delta \omega _{\mu}} +|0\rangle \langle B|(\hat{b}_{\mu}^{\dagger})e^{i \Delta \omega _{\mu}})d\mu \nonumber
\end{eqnarray}
where $\Delta \omega _{\mu} = \omega_{\mu} - D_0-g\mu _bB$ and $|B\rangle =\frac{1}{\sqrt{2}}(|1\rangle +|-1\rangle )$. 
Here, we can consider a subspace spanned by $|B\rangle $ and $|0\rangle $
because
the dark state $|D\rangle=\frac{1}{\sqrt{2}}(|1\rangle -|-1\rangle ) $ is not involved in the dynamics,
In this case, we can adopt the standard results of the open quantum system for a qubit. We consider the following Lorentzian form factor
\cite{koshino2005quantum}.
\begin{eqnarray}
 |g_{\mu}|^2=\frac{\gamma}{2\pi}\frac{\Delta ^2}{(\mu -\mu _0)^2 + \Delta ^2} 
\end{eqnarray}
where $\gamma$ denotes the strength of the noise, $\mu _0$ denotes a central frequency of the environment, and $\Delta$ denotes the linewidth of the form factor. In our paper, we set $\mu _0 = D_0 \simeq D_0 +g \mu _b B$ where $g \mu _b B$ is assumed to be much smaller than any other parameters.
We consider an initial state as $|B\rangle$, and we can calculate the survival probability as follows \cite{koshino2005quantum}:
\begin{eqnarray}
 P_{\rm{suv}}&=&|f(t)|^2 \nonumber \\
 f(t)&=&\frac{1+\sqrt{1-2\gamma /\Delta}
 }{2\sqrt{1-2\gamma /\Delta}}e^{-i \lambda _1 t}
 -
 \frac{1-\sqrt{1-2\gamma /\Delta}
 }{2\sqrt{1-2\gamma /\Delta}}e^{-i \lambda _2 t}
 \nonumber \\
\end{eqnarray}
Also, $\lambda _1$ and $\lambda _2$ are the solutions of the following equation:
\begin{eqnarray}
\lambda (\lambda +i\Delta) - \gamma \Delta/2=0 .
\end{eqnarray}
The density matrix in the interaction picture
after a time $t$ is described as follows:
\begin{eqnarray}
\rho_I(t)= P_{\rm{suv}}|B\rangle \langle B| + (1-P_{\rm{suv}})|0\rangle \langle 0| .
\end{eqnarray}
Therefore, by going back to the Schr\"{o}dinger picture,
we have 
\begin{eqnarray}
\rho(t)= P_{\rm{suv}}|\psi (t)\rangle \langle \psi (t)| + (1-P_{\rm{suv}})|0\rangle \langle 0|,
\end{eqnarray}
where $|\psi (t)\rangle = \frac{1}{\sqrt{2}}(|+1\rangle +e^{i\omega t}|-1\rangle )
$ and $\omega =2g \mu _bB$.
We can thus
define an error rate as $\epsilon =1- P_{\rm{suv}}$.

\bibliographystyle{apsrev4-1}
%

\end{document}